# Survey of Rigid Body Simulation with Extended Position Based Dynamics

Miguel Luis Nunes Seabra, Daniel Simões Lopes, João Madeiras Pereira
*Instituto Superior Técnico*
*Universidade de Lisboa*
Lisboa, Portugal
miguel.l.n.seabra@tecnico.ulisboa.pt, daniel.lopes@inesc-id.pt; jap@inesc-id.pt

*Abstract*—Interactive real-time rigid body simulation is a crucial tool in any modern game engine or 3D authoring tool. The quest for fast, robust and accurate simulations is ever evolving. PBRBD (Position Based Rigid Body Dynamics), a recent expansion of PBD (Position Based Dynamics), is a novel approach for this issue. This work aims at providing a comprehensible state-of-the art comparison between Position Based methods and other real-time simulation methods used for rigid body dynamics using a custom 3D physics engine for benchmarking and comparing PBRBD (Position Based Rigid Body Dynamics), and some variants, with state-of-the-art simulators commonly used in the gaming industry, PhysX and Havok. Showing that PBRBD can produce simulations that are accurate and stable, excelling at maintaining stable energy levels, and allowing for a variety of constraints, but is limited in its handling of stable stacks of rigid bodies.

## I. INTRODUCTION

### A. Motivation

Physical Simulation is a wide field within computer graphics and animation, being crucial for modern animation effects and interactive simulations such as those found in video games. The demand for reliable physical simulation has grown with the popularization of virtual reality since users interact with everyday objects within simulated environments in more ways than ever before. Physical simulators are usually measured along three metrics: robustness, accuracy, and time efficiency. Scientific simulations usually trade efficiency for accuracy, but for interactive applications, prioritizing fast solutions is key. Rather than true physical accuracy, real time simulations just need enough to maintain visual plausibility. Most of the time this means finding approximate solutions as fast as possible in a robust way.

A large body of work exists on how to speed up simulations and finding solutions for a vast range of physical phenomena and materials [2], [3], [4], [15]. PBD (Position Based Dynamics) is one of these methods, and it stands out from the crowd due to its 10.1109/TPS.2010.2090905, visual accuracy, and speed [4], as well as its ability to handle over and under constrained environments gracefully [14]. Its most common applications are the simulation of particle systems which are unable to efficiently simulate rigid bodies. Due to this limitation particle systems, and rigid bodies are often simulated in different physics engines meaning that interaction between both will be heavily limited.

Recently, PBD was expanded upon, creating Position Based Rigid Body Dynamics (PBRBD) which allows rigid bodies and particles to coexist and interact implicitly while remaining fast, robust, and stable. This method was first shown alongside a collection of demos showcasing its capabilities and strengths. These demos however did not contain any comparison to other simulators available [14]. Making it hard to tell how this method compares in terms of accuracy, speed and robustness to any current methods.

### B. Contributions

The work will focus on exploring the limits and benchmarking and comparing PBRBD to other methods for rigid body simulation helping future researchers and industry professionals to assess whether they should implement or perform further testing using PBRBD.

To benchmark, PBRBD a package was developed containing an implementation of the physics engine that can be used within the unity game engine. Benchmarking software was also developed that can create equivalent scenarios across three different independent physics engines, the custom PBRBD implementation, Unity's default physics system PhysX and Havok. This software also acquires benchmarking data for the three engines.

In short, this work will:
- Detailed the implementation of PBRBD in the Unity game engine.
- Present and discuss comparisons between PBRBD, PhysX and Havok.
- Present conclusions regarding PBRBD´s strengths and shortcomings.

## II. RELATED WORK

Rigid body dynamics simulation methods fall under three general categories, force, velocity and position-based methods [3]. Force methods solve collisions using virtual springs to enact forces on bodies and maintain any constraints. Since virtual springs act as natural forces some consider these methods to be more realistic. [5]. The most common types of force-based method are penalty methods. When a collision is detected a spring is created at the contact point that pushes the bodies into non-colliding positions [11]. To simulate friction a spring is created between the contact point that opposes tangential movement [17]. Common issues with this methodology are that resting contacts may suffer from oscillations from the springs [6], and collisions between fast moving or heavy bodies require strong springs to separate them which can lead to numeric instability [6], [7].

Velocity based methods solve collisions by changing velocities directly, Impulse methods do so via the application of impulses which are instantaneous accelerations acting during a
single instant [10], [16]. These methods might require resting contacts to be handled differently as the impulses used for separating objects can lead to jittering.

Position based methods were originally used for particle systems. These methods work on positions directly projecting objects currently colliding into the nearest collision free position using a Gauss-Seidel step to iterate and solve all collisions and constraints. The original Position Based Dynamics method [12], although fast, robust and simple, had shortcomings that made it harder to work with. The stiffness of constraints was time step dependent. This made arriving at a suitable stiffness parameter and sub step count a
challenge. The algorithm also had no direct correspondence to real world elastic and dissipation energy potentials [5], [8] making it hard to simulate real world scenarios. The original method's shortcomings were eventually solved by XPBD (Extended Position Based Dynamics) [8]. Currently, the term PBD is usually interchangeable with XPBD. The extension managed to decouple stiffness from substep count by replacing the concept with compliance, the inverse of stiffness. It also made the method more robust at handling hard constraints since they were essentially infinitely stiff. Using compliance, a hard constraint has a compliance of value one, and a fully compliant constraint has a value of zero [8]. It was also extended to receive physical quantities and introduced Lagrange multipliers to its equations, which offers the previously mentioned constraints a force estimate value. This method still had some shortcomings, it is derived from an implicit time stepping scheme and as such suffers from energy dissipation [8]. It has also been stated that Gauss-Seidel solvers can oscillate between solutions rather than converge given non-feasible sets [5]. The most recent iteration of PBD extends the method to handle rigid body dynamics. PBRBD (Position Based Rigid Body Dynamics) adds steps to XPBD's algorithm in order to handle orientation and angular velocity, as well as adding angular constraints [14]. It has also been shown that for position-based methods it is more efficient and accurate to break down the temporal window of each simulation step and simulate more steps per second handling, even at the expense of only performing one Gauss-Seidel iteration [9].

## III. IMPLEMENTATION

### A. Environment

To test and benchmark the algorithm an environment was needed that provided several physics engines to serve as comparison, and preferably also offered rendering, profiling and debugging tools. Unity's game engine using C# scripts to run the simulation was deemed the better option. Since PBRBD requires double precision floating points and most math capabilities provided by the engine only support single precision floating points none of Unity's math classes containing vectors, quaternions and matrixes could be used creating a fully independent package that handles all the simulation and then simply updates the positions Unity uses.

### B. Simulation loop

---

**Algorithm 1** Simulation Step

*BroadCollisionDetection*()
$h \leftarrow \Delta time/numSubsteps$
**for** *numSubsteps* **do**
    *PositionalUpdate*()
    *ConstraintSolve*()
    *VelocityUpdate*()
    *NarrowCollisionDetection*()
    *VelocitySolve*()
**end for**
*UpdateEnginePositions*()

---

The implemented simulation loop is executed once per frame and simulates as many substeps as *numSubsteps*. Increasing this value decreases the size of the temporal window, which has been shown to be the most efficient way of increasing accuracy [9]. The rigid bodies and particles used by the simulation are independent entities from the ones used by the game engine for rendering. As such, one final step is necessary to update the positions and orientations of the simulated bodies.

### C. Bodies

Particles are defined as a mass, position and velocity. A rigid body on the other hand also has orientation defined as a quaternion and angular velocity and external torque parameters defined as vectors. In order to properly apply realistic rotations, rigid bodies also require an inertia tensor which refers to mass in rotational terms [14], a 3x3 matrix that contains information regarding the moment of inertia of a rotation along the bodies' principal axes. Since shape and consequent mass distribution of a body has an impact on its rotation. The Inertia tensor is dependent on the orientation of the body, to avoid recalculation the tensor is always defined in local coordinates and any rotation that is applied to the body need to be converted to self-coordinates, multiplied by the tensor and converted back to world coordinates.

### D. Positional Update

The first step within the algorithm's internal loop performs the time integration of the current positions and velocities according to the current velocity and acceleration. During this step the previous position and orientations are updated. The

new position and velocity of a body is updated by applying one Euler step:

$$x = x + v * h \quad (1)$$
$$v = v + F * h \quad (2)$$

Where $x$, $y$ and $F$ are the position, velocity and external force vectors respectively, and $h$ is the time interval being simulated. The above update is sufficient for simulating particles, for rigid bodies the following steps are also required:

$$q = q + 0.5 * [0, w.x, w.y, w.z] * q * h$$
$$q = |q| \quad (3)$$
$$w = w + h * I^{-1} * (T - (w \times (I * w))) \quad (4)$$

Where $q$ and $w$ and $T$ are the orientation, angular velocity in self coordinates and external torque vectors respectively, $I$ represents the inertia tensor. Angular velocity is converted into a quaternion and transformed into world coordinates and scaled by the time $h$. Note that since the angular velocity $w$ is stored in self coordinates, there is no need to perform any coordinate conversion before applying the tensor.

*E. Constraints*

It is possible to create constraints that simulate a variety of physical effects, the Nonlinear Gauss-Seidel solver is capable of processing different types of constraints since all are solved in the same generalized manner, with positional and angular constraints requiring slightly different approaches.

Constraints vary in how the error and its gradient $\Delta C$ is calculated. The gradient is a vector that points in the direction with most impact to the error value and with magnitude proportional to the impact moving the object will have on the error value calculated by $C(x)$. The Lagrange multiplier used to solve the constraint by calculating the positional correction $\Delta x$ is calculated using the error value and the inverse mass values $w_i$ of bodies affected by it.

$$\Delta x = \lambda w_i \Delta C \quad (5)$$

$$\lambda = -\Sigma \frac{C(x)}{w_i \Delta C_i^2 + a/h^2} \quad (6)$$

Constraints use the value of inverse mass to distribute the correction between constrained bodies, as seen in 6 and 5. In practice, a body with twice the mass will suffer half the effect of the constraint's correction, while the lighter body will experience double. Using inverse mass is useful for having infinitely heavy objects that cannot be moved by any correction simply by setting its value to zero. Compliance $a$ determines how rigid a constraint should act. A compliance value of zero corresponds to a rigid constraint where error is fully corrected. More compliant constraints only correct a fraction of the error, leading to a spring like behavior.

For positional constraints, the Lagrange multiplier calculations are as shown in 6 and applied as in 5. Some positional constraints might not act on the center of mass of the body, in those cases a vector $r$ determines the offset from the acted on position and the center of mass, in self coordinates. Furthermore, the movement needs to impact the rotation of the body $q$. This is achieved via an extra step, correcting orientation:

$$\Delta x = q^{-1} * \Delta x$$
$$rotation = q * (I^{-1} * r \times \Delta x)$$
$$q = q + 0.5 * rotation * q \quad (7)$$
$$q = |q|$$

In order to keep energy conservation when transferring positional kinetic energy to rotational kinetic energy, a different value for inverse mass is used called the generalized inverse mass $W_i$.

$$rotation = r \times (q^{-1} * \Delta x)$$
$$W_i = w_i + (rotation * I^{-1}) * rotation \quad (8)$$

Distance constraints take two bodies, and offsets from the center of mass, and ensure that the distance between the positions with offset applied are within a certain range. The error of a distance constraint is simply the difference between the real distance between the points and the desired distance. The gradient points in the direction opposite to the other particle. And the magnitude of the gradient has a magnitude of one.

Angular constraints are solved similarly to positional ones. Instead of using inverse mass, the inverse inertia tensor, the rotational equivalent of mass, is used. Corrections come in the form of a vector which can be broken down into length $\theta$ and direction $n$. The new generalized inverse mass used for calculating the Lagrange multiplier is calculated as:

$$n_{self} = q^{-1} * n$$
$$W_i = n_{self}^T I^{-1} n_{self} \quad (9)$$

The correction of the orientations, with respect to the Lagrange multiplier $\lambda$ is done as follows:

$$p = n * \lambda$$
$$p_{self} = q^{-1} * p$$
$$p = q * (I^{-1} * p_{self}) \quad (10)$$
$$q = q \pm 0.5 * [p.x, p.y, p.z, 0] * q$$

Angular constraints are applied in relation to some axis, defined in the body's self coordinates, labeled $a_n$. In certain cases a secondary axis is needed, which takes the form of $b_n$. A hinge joint works by ensuring that two axes belonging to two bodies remain aligned. The gradient and error of this constraint can be calculated as:

$$\Delta C = \frac{a_{1\,world} \times a_{2\,world}}{|a_{1\,world} \times a_{2\,world}|}$$
$$error = |a_{1\,world} \times a_{2\,world}| \quad (11)$$

Ball joints work by limiting the angle between two axes to be in a certain interval. The angle between two axes ($\sigma$) is calculated and only if it exceeds the max bound ($a$) is an

error returned. The gradient is the same as a hinge joint and the error of the constraint is calculated as:

$$error = |a_{1world} \times a_{2world}| * (\sigma - a) \quad (12)$$

### F. Collisions

In PBD collisions are handled as constraints, when a collision is found a new constraint is initialized and corrected immediately. The gradient of this constraint is the vector that can separate the penetrating colliders in the shortest distance coinciding with the contact normal multiplied by the penetration depth at the contact point. To simulate correct restitution and friction, a special step, called the velocity solve, is needed, iterating through collisions, and adjusting velocities. When a collision is detected a collision data structure containing references to both colliders, collision point ($p$), penetration distance ($d$), and contact normal ($n$) is created.

### G. Restitution

To achieve physically accurate conservation of momentum, the velocities resulting from a collision in respect to the bodies' restitution coefficients $e_n$ need to be adjusted during the velocity solve step. A body with a restitution coefficient of zero absorbs all the energy from a collision impulse, while one with a value of one absorbs no energy.

This step handles collision instances that have a collision normal $n$, both colliders and both $r_n$ a value referring to the offset from the colliders center of mass to the collision point, in self coordinates. The velocity solve step begins by calculating the difference between both velocities $\Delta v$, the velocity normals $v_n$ and tangential velocities $v_t$.

$$\Delta v = n(-v_n + min(-(e_1 e_2)v'_n)) \quad (13)$$

This step consist of subtracting the current velocity and replacing it with a reflected velocity $v'_n$ [14]. The resulting correction to velocity $\Delta v$ now needs to be distributed by both bodies according to their masses and distributed in terms of positional and rotational energy. This following step is used whenever a velocity is corrected within the velocity solve:

$$\begin{aligned} p &= \Delta v/(w_1 + w_2) \\ v_1 &= v_1 + p/m_1 \\ v_2 &= v_2 - p/m_2 \\ w_1 &= w_1 + I^{-1}(r \times p) \\ w_2 &= w_2 - I^{-1}(r \times p) \end{aligned} \quad (14)$$

### H. Friction

Friction is a dissipative force that opposes movement between two tangential surfaces. The strength of this force is determined by the amount of force the bodies are exerting on each other (usually the normal force) and the friction coefficients, values referring to the amount of friction produced by the body's material. There are two types of friction, one that acts when initiating motion (static friction) and another that acts after movement is initiated (dynamic friction).

### I. Static Friction

Static friction is implemented using a positional constraint initialized after separating the contact. It takes the sliding bodies' positions $x_n$, the offsets from center of mass to collision points $r_n$ and the collision normal used for calculating the collision tangent direction. It then ensures that no tangential movement occurs between the contact points. The force exerted by a constraint can be calculated as:

$$\begin{aligned} F &= \lambda n/h^2 \\ \tau &= \lambda n/h^2 \end{aligned} \quad (15)$$

The formula that determines if static friction is applied in respect to the static friction coefficient $\mu_s$ and the normal and friction forces is:

$$F_{static} <= \mu_s F_{normal} \quad (16)$$

The values of $F_{static}$ and $F_{normal}$ are proportional to $\lambda_{static}$ and $\lambda_{normal}$ respectively, as such the formula above can be implemented as follows:

$$\lambda_{static} <= \mu_s \lambda_{normal} \quad (17)$$

If the above condition is not met then the constraint responsible for applying static friction is discarded before applying corrections to any bodies and marks the collision for dynamic friction to be applied during the velocity Solve step.

### J. Dynamic Friction

During the velocity solve step, collisions that have not experienced static friction have dynamic friction applied. The force exerted cannot exceed a certain value, determined by the dynamic friction coefficient and normal force. Using (15) and (1) it is possible to calculate the velocity the dynamic force produces during the current time step using the following formula:

$$\Delta v = \left| \frac{\mu_{dynamic} * \lambda_{dynamic}}{h} \right| \quad (18)$$

### K. Interaction

For behaviours that require input from the user such as using the mouse to drag bodies it is necessary to take into account that the input devices are updated at a lesser frequency that the simulations substeps, as such mouse positions need to be collected and interpolated by each substep to simulate continuous movement of the anchor point.

### L. Soft Bodies

Soft bodies are simulated as a series of particles connected by constraints, a mesh is then dynamically altered so that it matches the particles' positions [13]. Particles are connected via distance and also volume constrains which take four particles and ensure that the tetrahedron formed by their points' volume remains constant. The direction of the gradient is different for all particles, but it is always perpendicular to the plane defined by the three other particles.

## M. Jacobi Solver

In order to compare the difference between using a Jacobi and the default Gauss-Seidel solver, both solvers where implemented with an option to toggle between both . When using the Jacobi solver corrections are stored, once a substep is finished all the corrections are averaged and then applied to the body.

## N. Optimizations

Collision detection can be broken down into the broad and narrow phases. The broad phase takes all colliders and tries to identify which pairs of collisions can possibly occur during the next simulation loop and is executed once per step. The narrow phase iterates through the likely collisions and checks for actual contacts and is executed once per substep. The collision detection and constraint resolution steps are implemented in parallel, distributing the workload amongst different threads. For the Gauss Seidel solver, mutual exclusion blocks are used to ensure that no constraint that shares a body with another is processed in parallel.

## IV. RESULTS

### A. Testing Methodology

In order to properly benchmark and compare PBRBD's custom implementation, which will be referred to simply as PBD, with other prominent physics engines available in Unity, the default implementation using PhysX which will be referred simply as Unity and Havok's physics package, as well as a version of PBD using a Jacobi solver labelled as Jacobi and parallel configurations of both versions. A variety of topics of interest were selected, testing the engine's along an array of different attributes. Each topic is then be studied by proxy of simulation scenarios designed to expose each engine's performance in each topic using instrumentalized versions of the algorithms and data collectors. All tests were be conducted on a Lenovo Legion 5 laptop, using an AMD Ryzen 7 5800H processor, 16GB of RAM, and a GeForce RTX 3060 GPU.

### B. Momentum conservation in collisions

Conservation of momentum dictates how real world object's velocities are affected by a collision, ensuring that no energy is gained or lost from the collision's impulses. A commonly used mechanism for demonstrating the phenomenon is Newton's cradle a device that, excluding dissipative forces can remain in perpetual motion, comprising, usually, of four spheres of equal mass suspended by wires. Simulating the contraptions using distance constraints to simulate the strings as in Fig. 1 shows that PBD provides the closest simulation to theoretical results as supported by the results in Fig. 2.

### C. Energy conservation in constraints

On the above section, the analyzed energy losses were, mostly, a byproduct of several collisions in a short time frame. While the triple pendulum (Fig.3) is a suitable test case to for the accuracy and stability of distance constraints it does not test the impact that its corrections had on orientation. In order to test a scenario where proper simulation of the orientation of bodies is crucial, a chain was simulated by attaching the edges of capsules together. At the end of the chain, a heavy sphere is attached (Fig. 5) in order to weigh the chain down. According to Fig. 6 PBD offers the best results for simulations of 100 capsules, displaying negligible energy gains and acceptable losses. However, when simulating 500 capsules, the simulation breaks and massive amounts of energy are gained. This most likely has to do with the velocity recalculation step, after a large correction was applied to a particle its velocity was recalculated as a massive value leading to unexpected and uncontrollable behavior.

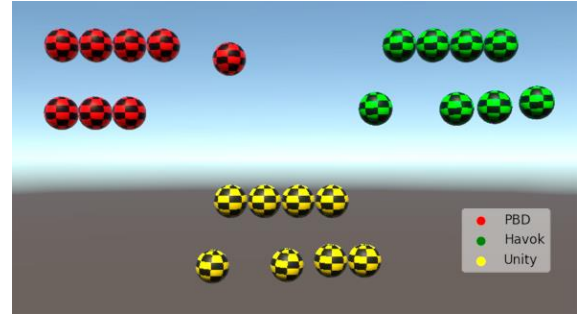

Fig. 1: Result of the first collision of the simulated Newton's cradle, with Unity and Havok's simulations showing wrongful momentum transfer as no velocity should be transmitted to the two middle spheres. The spheres on the top row of each simulation are static and are connected to the bottom layer via distance constraints.

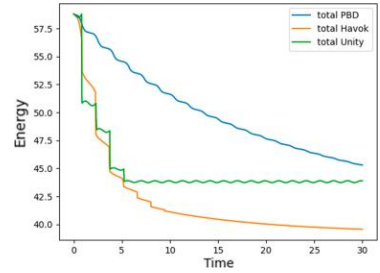

Fig. 2: Total energy values of Newton's cradle using constraints to simulate string over 30 seconds. PBD shows a stable energy loss while Havok and Unity show sudden drops that coincide with the times collisions happen during the simulation. The reason their energy levels stabilize is because the spheres at the center are keeping some velocity and swinging, the simulation converges to a state where all four spheres swing in tandem.

### D. Stability

Stacks of bodies are challenging simulations because error can quickly propagate and bring, what should in theory be a stable stack, to collapse in on itself. To test the stability of a stack, not only will energy conservation be analyzed but also how much the body at the top of a stack, which should be stationary, moves(Fig. 8) in order to capture positional error in the form of drift, a known problem with velocity methods [8]. How much the top element of the stack moved can then be analyzed by looking at the difference between the bodies starting and current positions throughout the simulation. In order to distinguish between error concerning the separation of penetrating bodies and drift, horizontal and vertical deviation from the starting position is analyzed separately. The most basic example of the type of system mentioned above would be a single vertical stack of cubes(Fig. 7), where the forces experienced by cubes near the bottom of the stack would be immense due to having to support hundreds of other cubes on top of them. There are many ways positional error can impact said structure. The stack may become compressed as the system is unable to keep bodies from inter-penetrating, efforts to separate bottom cubes may increase penetration towards the top which may end up with cubes being sent upwards instead of resting, PBD makes no distinction between resting contacts and collisions making it more susceptible for inter-penetration to occur. A more complex type of stack would be to organize the cubes in a pyramid shape(Fig. 9). PBD suffers from rotation being applied to cubes, this happens because each cube has four contacts beneath it, using the Gauss-Seidel solver means that each contact is solved and corrected one at a time, when the first contact is handles, towards one of the bottom corners of the object it applies a rotation as well as a translation, this rotation will cause further penetration in the opposite corner which eventually leads to an overcorrection and destabilizes the stack. When using Jacobi the corrections are stored and averaged together and applied all at once meaning that the rotations will cancel each other out leading to a more stable stack. Jacobi also suffers from some vertical oscillations, similar to what was shown in the previous example. Neither method achieves a stable stack at every simulation, but Jacobi has a higher rate of success. Unity's simulation is vertically and rotationally stable but suffers from horizontal drift, when the cubes are too spread out the structure collapses. Finally, Havok provides a stable simulation and could still provide stable results with more than five times as many elements being the strictly better choice.

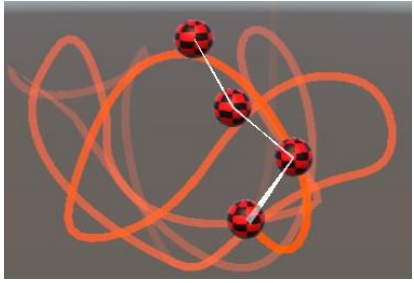

Fig. 3: Triple pendulum simulated using four particles and distance constraints.

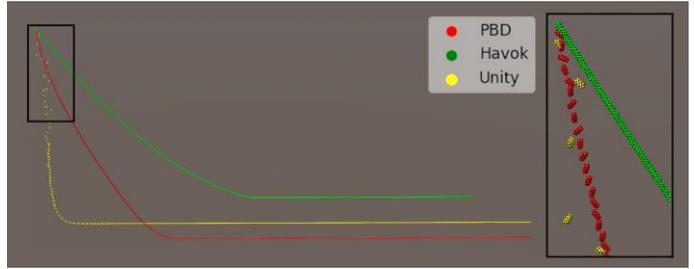

Fig. 5: The simulation of a chain made up of 500 capsules connected by constraints. Both PBD and Unity exhibiting wrongful simulation towards the top of the rope, with the latter showing the most error.

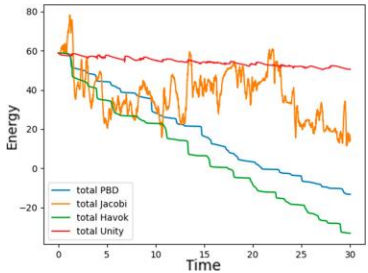

Fig. 4: Energy evolution over a 10-second simulation of a triple pendulum using distance constraints. Unity provides greater energy conservation, however it does suffer from some energy being gained which might compromise its simulation in certain scenarios. Unity and Havok exclusively lose energy and show similar results. Meanwhile Jacobi is very unstable in comparison suggesting the solver is less accurate.

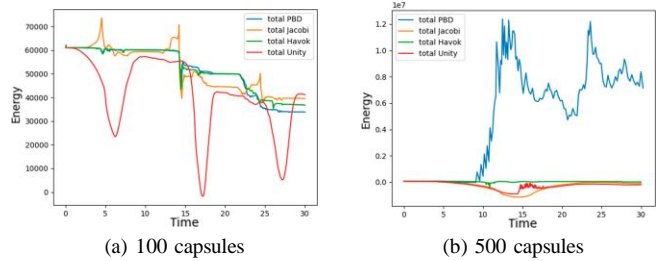

(a) 100 capsules      (b) 500 capsules

Fig. 6: Energy conservation when simulating a chain comprised of a varying number of capsules connected by constraints. All engines used 20 iteration or substeps. While PBD is shown to be the most stable choice for 100 capsules its also shown that when a scenario is too complex for the current substep count it can diverge. Jacobi which has been shown to be is less accurate in Fig. 4 is shown here to be more robust.

### E. Linear and Rotational Kinetic Energy

When calculating the energy values of a system, two separate values make up the total energy value, potential energy is dependent on the height of an object, and kinetic energy on its velocity. Kinetic energy has two components as well, linear kinetic energy and rotational kinetic energy. Following a

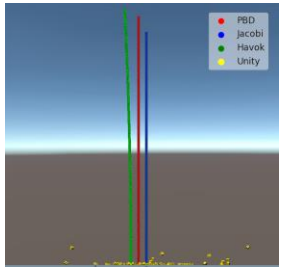

Fig. 7: Simulation of 100 stacked cubes. Havok's simulation is suffering from drift, tilting to the right and about to collapse, Unity has already fallen due to the same issue, meanwhile PBD and Jacobi cannot fully stop cubes from inter-penetrating leading to them having less height than they should and oscillating.

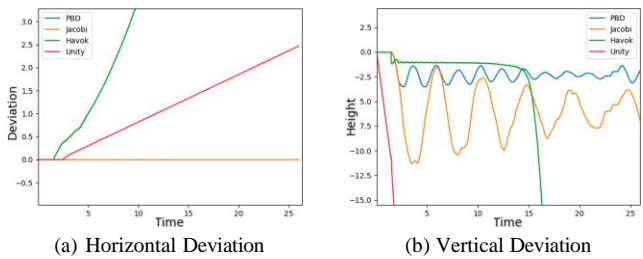

(a) Horizontal Deviation    (b) Vertical Deviation

Fig. 8: The horizontal and vertical deviation from the top cube's current and starting position over time in a vertical stack. Havok's and Unity's simulation eventually collapses, seen by the large vertical deviation. PBD and Jacobi show no drift, but they both oscillate vertically as they struggle to keep the cubes from interpenetration, at worse, this could result in cubes gaining vertical velocity and being thrown upwards.

collision, a falling object might transform potential energy into kinetic energy, and that energy needs to be properly distributed by its linear and rotational components. This distribution is a potential source of error for simulations. Errors in the distribution between both types of kinetic energy are more noticeable in objects with significantly different moments of inertia for each primary axis, as would be the case in a cuboid with different lengths for each axis, as seen in Fig. 10. PBD and Jacobi are both able to maintain stable energy levels but show some mild signs of energy being gained. Unity and Havok suffer from significant losses and show energy spikes as well, although Havok's can quickly fix errors due to the use of error caches. The errors present in both Unity and Havok's simulations seem to originate from too much rotation being applied. Most likely originating from having to solve the positional error and apply restitution by applying an impulse. Unity is the most unstable out of the tested engines since it shows energy rising continually, this error is visually apparent was well as the plane gradually reaches greater heights. Havok's shows some energy gain but still converges into a low energy state. This might be because

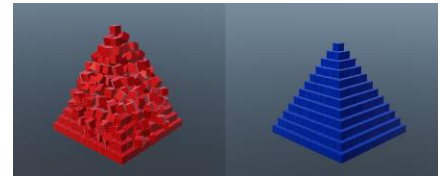

(a) PBD Simulation    (b) Jacobi Simulation

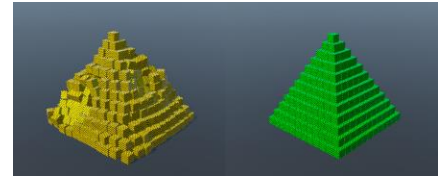

(c) Unity Simulation    (d) Havok Simulation

Fig. 9: The simulation of 650 cubes organized in a pyramid shape. PBD and Unity's simulation show wrongful behavior leading to the eventual collapse of the structure, the former's error comes from rotations within the bodies while the latter's is due to drifting.

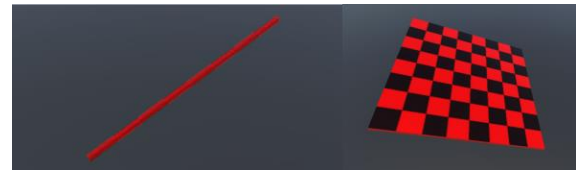

(a) Rod Cuboid    (b) Plane Cuboid

Fig. 10: Cuboids with uneven axes lengths

Havok detects ambiguous situations where energy gains due to rotation are likely and always minimizes energy [1].

*F. Friction*

In order to test each simulations implementation of friction cubes and spheres were placed in large cuboids serving as a steep plane (Fig. 12). In a scenario with a cube sliding down a ramp with frictional forces present, it is expected that energy is gradually lost. After the cube leaves the ramp and begins a free fall, energy should remain constant. Analyzing the energy in each cube system (Fig. 13a) reveals Havok's to be the only accurate one, where energy is lost continuously, visually the cube slides down the ramp as expected. PBD, Jacobi and Unity's simulations are not as accurate, having the cube gain a rotation and tumble down the ramp instead of sliding. This can be confirmed by analyzing the energy values, as energy is lost abruptly with each collision. The reason for Havok's handling of the scenario probably has to do with Havok being able to recognize situations where rotation due to friction is likely to be applied wrongfully, in these cases Havok always chooses the option that minimizes energy in the system [1], ignoring the rotation all together, this could also explain why Havok loses energy when the test is repeated with spheres (Fig. 13b), in that scenario PBD and Jacobi provide better results. It might also be the case that the difference might stem from the collision point used for the contact, if the point used is towards the edge of the cube it has more impact on the rotation then if it was centered in a face.

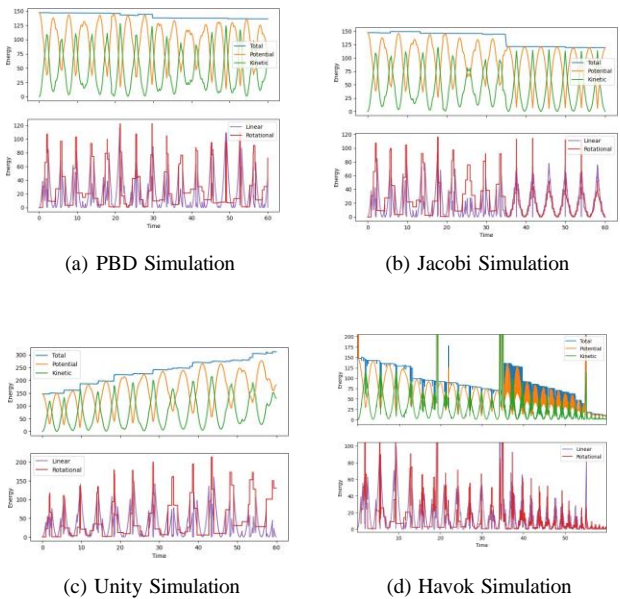

(a) PBD Simulation  (b) Jacobi Simulation

(c) Unity Simulation  (d) Havok Simulation

Fig. 11: The energy of the systems throughout a 30-second simulation of a plane cuboid.

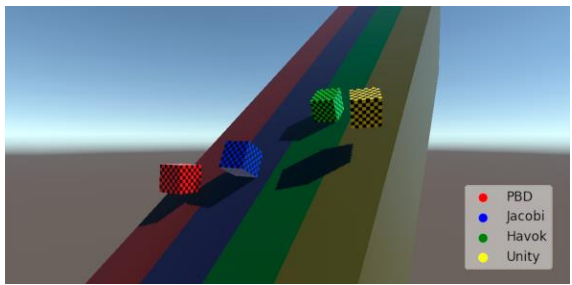

Fig. 12: Friction test with a cube sliding down a ramp, using a value of 0.3 for all friction coefficients.

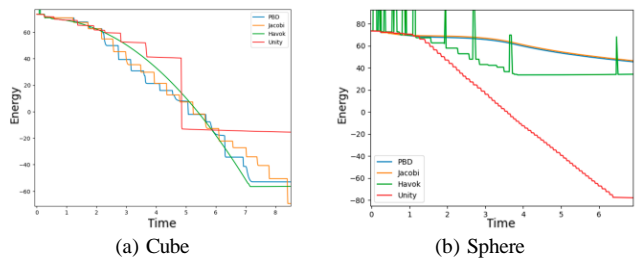

(a) Cube  (b) Sphere

Fig. 13: Energy of a rigid body sliding down a ramp over time.

### G. Physically Impossible Scenarios

Within the regular use of a game engine by a developer, or during the execution of a gameplay environment, it is likely that at some point a physically impossible situation arises, either from the developer setting up impossible starting conditions or the gameplay enforcing a specific state. It is an important factor when choosing a physics engine that it is capable of remaining stable in these situations while minimizing physical inaccuracy. While in this scenario, positional methods appear to be superior, further testing reveals that depending on the penetration depth and scenario they might get worse results. Repeating the same experiment but with a pyramid of penetration depth of 0.4 (nearly half of the cubes' height) shows similar results for Havok, Unity's top layers get a vertical velocity, PBD and Jacobi the pyramid is dismantled within the first simulation step with cubes belonging to the base and the middle layer all at the same height. This happens because while the bottom layer is being processed its members are projected vertically in order to correct the penetration with the plane, however this correction puts them in a position where they penetrate cubes in the middle layer. For lesser starting penetration depths this causes no issues, but with higher starting penetrations the resulting intermediate state where the bottom and middle layer collision have so much penetration in the vertical axis that the shortest correction distance between each colliding cube is horizontal, leading to both layers expanding to each side, and dismantling the structure.

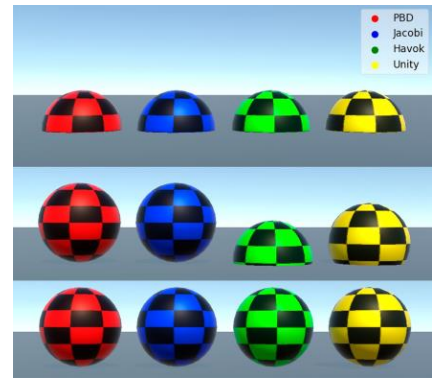

Fig. 14: Collision solve of a scenario with colliding starting positions. The velocity based methods used by Unity and Havok cause the collision to be resolved over several frames rather than immediately.

### H. Unsolvable Constrained Scenarios

In any engine dealing with constrained systems, it is possible to create a configuration that is unsolvable due to having constraints with mutually exclusive solution domains. As was tested in Fig. 15. Havok provides a stable approximation of a state that averages each constraint to minimize error, the chain is still straight, and each link remains static, being the best option for this scenario. Jacobi manages a stable and static simulation as well, but the state at which it arrives is does not fully minimize error, as there is a slight angle to each link. Finally, Unity and PBD oscillate in their solution with Unity even showing signs of divergence, with capsules moving to seemingly random positions before stabilizing at an oscillatory state switching between both positions at each step. This is a known issue with the Gauss-Seidel solver which Jacobi avoids, and it can lead to behavior that is visually striking. The issue can be mediated by adding some compliance to each constraint, in which case the system converges to a stable and error minimizing state.

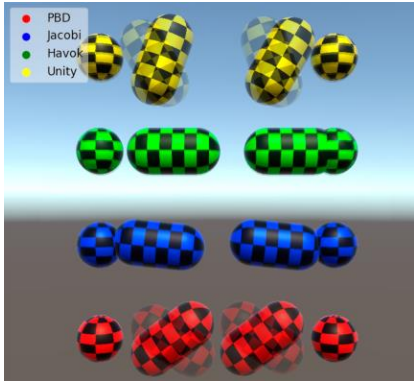

Fig. 15: A chain with each end attached to a static point whose distances are greater than the chains' length, leading to a scenario with incompatible constraints. Havok and Jacobi can converge at a stable state, while Unity and PBD oscillate. The image consists of two over imposed different screenshots.

*I. Performance*

In order to test how the performance of each engine scales, three scenarios were tested with increasing number of elements, the chain (Fig. 5) to test the impact constraints have on performance, the pyramid (Fig. 9), mean to test the performance impact of resting contacts, and a new scenario consisting of a pile of capsules which will be referred to simply as capsule to test the impact of an unstable stack. Since it is possible to optimize resting stable contacts, this test might produce significantly different results than a stable stack. Note that both Havok and Unity's physics engine are implemented in C++ and most likely use GPU acceleration techniques as well, meanwhile our custom PBD implementation is in C# and also has the overhead of being processed as part of a Unity MonoBehaviour. It also only uses CPU threads, meaning that it is at a disadvantage, measuring the performance of the different engines is further complicated by the fact the biggest bottleneck in most physics engine remains collision detection, and this work is mostly focused on measuring the performance of collision restitution, time integration and constraint solving. While PBD and Jacobi's simulation seem to have similar performance to Unity when simulating constraints the implemented collision detection is not able to compete with the one provided by Unity and Havok. The most important factor in this analysis is that all engines scale linearly, meaning that none is inherently more complex. The slope of each result will be largely impacted by low level optimizations, implementation language and parallelization and not so much by the simulation method itself. When it comes to shear performance Havok is unmatched in its handling of large scenarios.

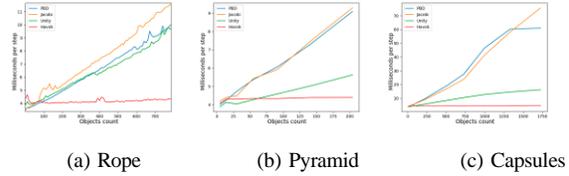

(a) Rope     (b) Pyramid     (c) Capsules

Fig. 16: Milliseconds per physics substep of simulations.

## V. DISCUSSION

After thoroughly testing all engines in different scenarios, it is clear no options is strictly better than the other, with each engine presenting unique strengths and issues. Table I combines the observations from all scenarios regarding problems with each method.

TABLE I: Physics Engine Comparisons

| | Engine | | | |
|---|---|---|---|---|
| | **PBD** | **Jacobi** | **Unity** | **Havok** |
| Velocity transfer | Accurate | Accurate | Inaccurate | Inaccurate |
| Constraint error | Noticeable | Noticeable | Very Noticeable | None |
| Inter-penetration | Significant | Very Significant | None | None |
| Velocity drift | None | None | Very Significant | Significant |
| Stacking | Rotational Error | Somewhat | Drifting | Stable |
| Rotational kinetic energy | Stable | Stable | Increase | Loss |
| Stable friction | No | No | No | Yes |

*A. Position Based Rigid Body Dynamics*

In terms of maintaining stable energy levels without producing any extra energy PBD is the most reliable option. Along with Jacobi it is the only method capable of accurately and quickly transferring velocity over a row of objects simulating a Newtons cradle with ease. It is stable handling of collision contacts is further supported by the results of (11a) being able to simulate objects with uneven axes lengths.

When dealing with constraints, where velocity needs to be recalculated as the difference in positions, it loses some energy gradually. It is, however, more stable, showing no energy being gained. It does come with the issue that when dealing with scenarios too complex for its current substep count (which dictates accuracy) the system will diverge causing massive energy gains. This is because PBD is meant to excel at correcting small errors, hence the use of substeping.

When it comes to scenarios with stacks of bodies, PBD can allow for some inter-penetration of bodies and suffer from rotational error destabilizing the stack.

*B. Jacobi Solver Position Based Rigid Body Dynamics*

Using a modified version of PBD to use a Jacobi solver rather than Gauss-Seidel proves to offer little benefit since the method cannot produce the same kind of stable energy. The issues of inter-penetration also present in PBD are more pronounced as well. However, in some cases it can be a better choice, when dealing with a scenario too complex for the current substep count PBD was shown to diverge, while Jacobi managed to maintain a more stable simulation as shown in (6b). Jacobi also suffers from less rotational error on resting stacks, but it is more vulnerable to destabilization due to the more severe interpenetration.

*C. Unity*

Unity's PhysX based physics engine's performance in the tests conducted revealed itself to be similar to Havok's. Some issues are shared by both engines, such as velocity transfer error. Both show velocity drifting, but Unity's was more severe. Unity is also prone to energy gains when tested using long cuboids (11c). The one advantage Unity was shown to have over Havok was better conservation of energy in constrained scenarios without collision, being able to maintain motion for a much longer time.

*D. Havok*

Havok's simulation proved itself to be incredibly stable and fast, being able to handle scenarios that were more complex and rarely showing any signs of divergence or overcorrection. However, it does suffer from velocity drift and loses a lot more energy than other engines. This is due to Havok detecting situations that could lead to energy being gained and always chooses the option that minimizes energy [1].

## VI. CONCLUSION

As the results show, PBRBD can simulate scenarios with great accuracy showing great conservation of momentum and no energy gains and is a solid choice for any game engine, not only due to its accuracy but also due to being a unified solution merging particle systems and rigid body dynamics in a single engine and allowing for many different types of constraints. Its shortcomings become more prevalent when simulating stacks, where the Gauss-Seidel solver struggles with rotational error. Switching the default solver by a Jacobi solver can increase robustness and handling of scenarios with mutually exclusive constraint solution domains but decreases the accuracy of the simulation. In short, this novel method will facilitate development of scenarios mixing particle system or destructible objects while providing accurate and stable results while avoiding some known issues with velocity-based methods such as velocity drifting.

## VII. FUTURE WORK

Implementing the algorithm in C++ with a focus on code optimization and GPU parallelization to understand how performant PBRBD can get could further establish it as a solid choice for rigid body dynamics. The algorithm requires double precision floating points, which GPUs are not optimized to handle, leading to possible challenges. Furthermore the parallelization of the Gauss-Seidel step can be achieved in a variety of ways, comparing the impacts on performance and accuracy of different techniques could be a source of future work. Due to its variety of supported constraints PBRBD also seems adequate to create controls such as buttons, cranks, levers and knobs commonly seen in VR applications which could be further studied with user tests.